\documentclass[letters,fleqn,usenatbib]{mnras}

\usepackage{newtxtext,newtxmath}
\usepackage{mathptmx}
\usepackage[T1]{fontenc}
\usepackage{ae,aecompl}
\usepackage{xcolor}

\usepackage{graphicx}	% Including figure files
\usepackage{amsmath}	% Advanced maths commands
\usepackage{amssymb}	% Extra maths symbols

\title[Molecular gas mass through the Balmer Decrement]{A new empirical method to estimate the molecular gas mass in galaxies.}

\author[Concas \& Popesso]{
Alice Concas,$^{1,2,3}$\thanks{E-mail: a.concas@mrao.cam.ac.uk (AC)}
Paola Popesso$^{1}$
\\
$^{1}$Excellence Cluster Universe, Boltzmannstr. 2  D-85748 Garching,  Germany \\
$^{2}$Cavendish Laboratory, University of Cambridge, 19 J. J. Thomson Ave., Cambridge CB3 0HE, UK \\
$^{3}$Kavli Institute for Cosmology, University of Cambridge, Madingley Road, Cambridge CB3 0HA, UK}

\date{Accepted XXX. Received YYY; in original form ZZZ}

\pubyear{2019}

\begin{document}
\label{firstpage}
\pagerange{\pageref{firstpage}--\pageref{lastpage}}
\maketitle

\begin{abstract}
We find a tight correlation between the dust extinction, traced by the Balmer Decrement (BD$=$H$\alpha$/H$\beta$), the CO(1-0) line luminosity (L$_{CO}$) and total molecular gas mass (M$_{H2}$) in a sample of $222$ local star-forming galaxies drawn from the xCOLD GASS survey.
As expected, the galaxy disk inclination affects the correlation by inducing a saturation of the Balmer decrement on highly inclined galaxies. Once this effect is taken into account, L$_{CO}$ and M$_{H2}$ can be expressed as a function of BD with a scatter of $\sim 0.3$ dex. We do not find any dependence on galaxy size, mass, morphology, star formation activity and gas metallicity. % to look for possible sources of scatter and bias.
The correlation disappears if the atomic gas phase is considered. This is likely due to the fact that the region traced by the BD, the stellar disk, is much smaller than the HI disk.
\end{abstract}


%%%%%%%%%%%%%%%%%%%%%%%%%%%%%%%%%%%%%%%%%%%%%%%%%%

%%%%%%%%%%%%%%%%% BODY OF PAPER %%%%%%%%%%%%%%%%%%

\section{Introduction}
%Dust and gas in galaxies are the key ingredients of star formation, which is governed by their complex, mutual interplay in the baryon cycle. 
The star formation activity of galaxies is governed by the complex relationship between gas and dust. While dust works as catalyst in transforming atomic hydrogen (HI) into molecular hydrogen (H2), the collapse of giant molecular clouds leads to the star formation process itself (e.g., \citealt{Wolfire1995}). It is clear, then, that the molecular gas is the most accurate tracer of the star formation regions and it is essential to understand the details of the process. 

{The molecular phase gas is not directly observable (see \citealt{KennicuttEvans2012} for a review). Commonly the luminosity due to the $J=1-0$ transition of the carbon monoxide molecule (CO) is used as a good proxy of the H2 gas mass (e.g.\citealp{KennicuttEvans2012, Bolatto2015, Saintonge+2011,Saintonge+2017}). However, despite the great potential of this method, CO observations are extremely time-consuming.  Also in era of  huge, new millimeter facilities, as the Atacama Large Millimeter/submillimeter Array (ALMA),  building a statistically significant CO galaxy survey is still challenging.}
%However, the direct measure of the molecular phase of the hydrogen is made very difficult by the lack of the dipole moment and the low mass of the H2 molecule (see \citealt{KennicuttEvans2012} for a review). To overcome this problem, commonly, two main approaches are used to unveil the "elusive" molecular gas reservoir on galaxies: indirect molecular tracers and calibrated scaling relations.

%The most used indirect indicator is the carbon monoxide molecule (CO), whose luminosity, L$_{CO}$, given by the $J=1-0$ transition, is believed to be a good proxy of the H2 gas mass (e.g.\citealp{KennicuttEvans2012, Bolatto2015, Saintonge+2011,Saintonge+2017}). However, despite the great potential of this method, the direct measurement of the molecular gas through CO observations is extremely time-consuming. Building a statistically significant CO galaxy survey is still challenging despite the advent of the new millimeter facilities as the Atacama Large Millimeter/submillimeter Array. 
%Indeed, the largest CO survey ever conducted is the xCOLD GASS survey (\citealp{Saintonge+2017}), which is limited to $\sim 500$ local galaxies.  

%implied that dust is a reliable tracer of cold gas.
A powerful alternative to estimate the molecular gas mass is the use of the well known empirical gas-to-dust mass relation (e.g \citealt{Bourne+2013,Scoville2014,Groves+2015,Bertemes+2018}). {The dust emission has been} used to estimate gas content in the Milky Way 
\citep[e.g.,][]{Dame2001}, and in nearby galaxies by a number of authors 
\citep[e.g.,][]{Israel1996,Israel1997,Israel2005,Leroy2009,Boquien2013}. In addition, it has been extended to sub-mm observations \citep{James2002} and more recently to Herschel data \citep[e.g.,][]{Eales2010,Eales2012,RomanDuval2010}. 
For instance, \cite{Scoville2014,Scoville2015,Scoville2016,Scoville2017} apply a similar concept to the Rayleigh-Jeans side of the spectral energy distribution of local and high-z galaxies. 
The cold gas mass is estimated by converting the luminosity at rest frame of 850 $\mu$m into cold gas mass with a scaling relation calibrated on local and high redshift galaxies (see \citealp{Scoville2016} for more details).

%The H2 column density can independently be inferred by measuring the optical or near-infrared light from background stars that has been extincted by the dust present in the molecular cloud (Lada et al. 1994; Cambre ́sy 1999; Dobashi et al. 2005). This method is often regarded as one of the most reliable because it does not depend strongly on the physical conditions of the dust. But this method is not without some uncertainty. Variations in the total to selective extinction and dust-to-gas ratio, particularly in denser clouds like those in Taurus, may introduce some uncertainty in the conversion of the infrared extinction to gas column density (Whittet et al. 2001). Dust emission has also been used to derive the column density

More recently, an alternative method has been proposed to retrieve the molecular gas mass based on the absorption rather than the emission of to the dust grains, (e.g. \citealp{Brinchmann2013,Kreckel+2013,Boquien+2013,Barrera-Ballesteros+2018}). Such methodologies are based on the fact that, in the HII regions, the young ionizing O stars are commonly associated with the clouds of cold gas from which they formed (\citealp{Calzetti+1994,CharlotFall2000}). 
{Such surrounding dust is able to absorb the energetic photons leaving an imprint in the UV and optical light, and leading to a well known correlation between the dust column density and the reddening or the extinction of starlight ( e.g. \citealt{KennicuttEvans2012,Kreckel+2013}).}
%Such surrounding dust is able to absorb the energetic photons leaving an imprint in the UV and optical light. {As a result, it is the well known correlation between the dust column density and the reddening or the extinction of starlight ( e.g. \citealt{KennicuttEvans2012,Kreckel+2013})}. 
{For this reason, the optical and infrared extinction has been used in several studies to inferred the molecular gas column density in our MW (see \citealp{Bohlin+1978,Lada+1994,Dobashi+2005,Pineda+2010}). 
In external galaxies, \cite{Brinchmann2013} proposed a theoretical approach to use the extinction of the optical emission line fluxes in SDSS local galaxy spectra to indirectly constrain the dust and gas masses of the HII regions. 
The main limitation of this method is the nature of the fitting procedure that requires the a priori knowledge of the extinction law models, making the results necessarily model dependent (see  \cite{Brinchmann2013} Section 3.2). }%\cite{Kreckel+2013} show also that the Balmer decrement (BD), the ratio of the Balmer line fluxes H$\alpha$ and H$\beta$, largely used to correct for dust extinction, is correlating with the dust mass surface density in the Herschel detected KINGFISH galaxies.

%To overcome in this problem,   with 
{In this Letter, we investigate with an empirical approach whether the optical attenuation by dust, as seen by reddening of the Balmer lines (H$\alpha$ and H$\beta$), can be used as a proxy of the molecular gas mass by taking advantage of the largest CO survey ever conducted, the xCOLD GASS survey (see \citealt{Saintonge+2011,Saintonge+2017}). }
In particular, we study the correlation between BD estimated from the SDSS galaxy spectra, L$_{CO}$ and M$_{H2}$, provided for CO detected local xCOLD GASS galaxies. 

In Section 2, we describe and characterize our galaxy sample. The L$_{CO}$-BD, M$_{H2}$-BD and M$_{HI}$-BD correlations are presented in Section 3. Finally, we summarize our main conclusions in Section 4. 
Throughout this Letter, the following cosmological parameters are assumed: $H_{0}=70$ km s$^{-1}$ Mpc$^{-1}$, $\Omega_{M}=0.3$ and $\Omega_{\Lambda}=0.7$.

% Singoli SN Hbeta 322 match galineDR7 e zCOLDGAS   in piano   %figura fatta con SFR_Mstar_plane_LCO_MH2_nuovo.pro SFR_Mstar_plane_LCO_MH2.pro IDL 8
   \begin{figure}  
   \centering
   \includegraphics[angle=0,width=\hsize ]{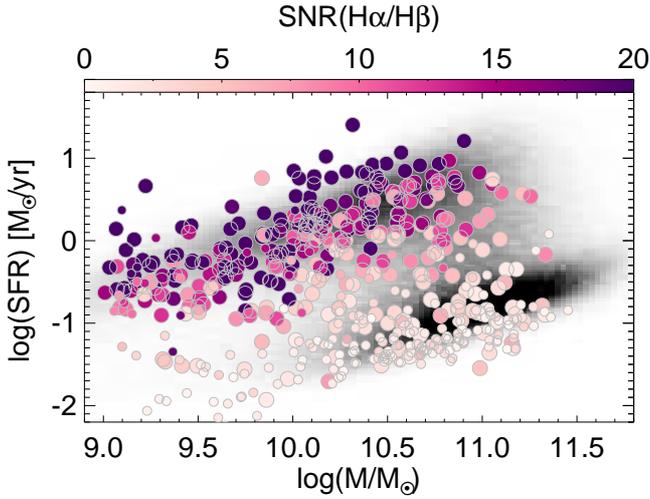}%{balmerDec_SFR_Mstar_plane_COobs_SNRBD_nuovo.eps}%{balmerDec_SFR_Mstar_plane_COobs_SNRBD.eps}%{balmerDec_MH2_SFR_Mstar_plane_COobs_SNRBD_+bertemes2018.eps}%{SFR_Mstar_plane_SNRBD_xCOLD_Bertemens+2018_seconda.eps}%{balmerDec_MH2_SFR_WISE_SED_Mstar_plane_tot.eps}
   %{density_DR7_VESPA_SFR_M_plane_MPAvalues2.ps}
   \caption[Distribution of our sample in the SFR-M${\star}$ diagram.]{Distribution of the xCOLD GASS galaxy sample (big and small circles, for the CO detentions and CO upper limits, respectively) in the SFR-M${\star}$ diagram, color-coded as a function of the Balmer Decrement signal$-$to$-$noise ratio, SNR(H$\alpha$/H$\beta$).
   The gray-scale contours show the overall SDSS population. Our sample perfectly covers the entire region of star-forming galaxies in a wide range of M$\star$.}
              \label{FigPlaneBalmerSingoli}%
   \end{figure}
%______________________________________________ 
\section{DATA}

\subsection{The CO molecular gas mass and Balmer Decrement}
The sample of galaxies analyzed in this work is taken from the extended CO Legacy Database for GASS survey (xCOLD GASS, \citealp{Saintonge+2017}), designed to provide a picture of molecular gas across the local galaxy population. The sample consist of $532$ galaxies observed at the IRAM 30 m telescope over the course of two large programs: the original COLD GASS survey (\citealt{Saintonge+2011}) targeted $366$ galaxies at $0.025 < z < 0.050$ with stellar mass M$\star > 10^{10}$ M$_{\odot}$, and the subsequent COLD GASS-low survey (\citealt{Saintonge+2017}), with $166$ galaxies with $ 10^9 <$ M$\star < 10^{10}$ M$_{\odot}$ in a similar redshift range (\citealt{Saintonge+2017}). %which extends the original sample down to stellar masses of $10^{9}$ $M_{\odot}$ in a similar redshift window.
%The galaxies were randomly selected from the SDSS parent sample to have a flat stellar mass distribution. To avoid this "mass-bias", a statistical weight is assigned to each object, obtained by comparing the mass distribution of the xCOLD GASS sample with the expected mass distribution of a purely volume-limited sample of $532$ galaxies based on the \cite{Baldry+2012} stellar mass function (see \citealt{Saintonge+2017} and \citealp{Catinella+2018} for more details).
%Any relation presented in this paper takes these weights into account, as they make the xCOLD GASS sample volume-limited and therefore representative of the local galaxy population with M$\star > 10^9$ M$_{\odot}$.
%About half of the sample ($532$ systems) has a CO(1-0) detection or upper limit in the IRAM-30m telescope observations. 
The galaxies with detected CO line are  $63 \%$ of the sample (333 objects), while for the remaining galaxies only  L$_{CO}$ and M$_{H2}$ upper limits are provided (see \citealp{Saintonge+2016,Saintonge+2017}).
%To increase the statistic of our sample we include in our analysis the recent 78 massive ($M{\star}>10^{10} M_{\odot}$) galaxies with known CO emission line flux presented in \cite{Bertemes+2018}. For all this galaxies we estimate the L$_{CO}$ following the same prescription showed in \cite{Saintonge+2017}.
%
% Singoli BDecrements Brinchmann correlazione LCO e MHII           %figura fatta con
%plot_correlation_BalmerDec_LuminositaCO_tutti_BPT.pro plot_correlation_BalmerDec_LuminositaCO_SF.pro plot_correlation_BalmerDec_LuminositaCO_tutti_NON_SF.pro
%plot_correlation_BalmerDec_Mmolecolare_COLD+bertemes.pro.pro + plot_correlation_BalmerDec_LuminositaCO_COLD+bertemes.pro.pro + knote plot_BalmerDecrement
   \begin{figure}    
   \centering
   \includegraphics[angle=-90,width=0.99\hsize ]{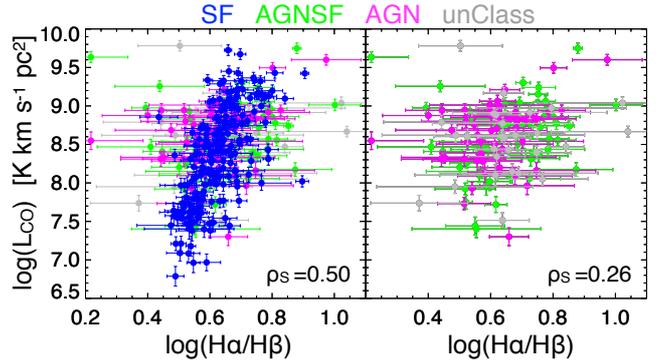}%{balmerDec_LumC0_correlation_XCOLDGASS_Cappellari_NON_SF.eps}%{balmerDec_LumC0_correlation_XCOLDGASS_Cappellari_all_BPT.eps}%{correlazione_Lvp_BD_allBPT.eps}%{Copiacorrelazione_3Plots_BPT.pdf}%{correlazione_3Plots_BPT.eps}%{BD_LCO_MH2_xCOLD_Bertemes+2018_SNR10.eps}%{BD_LCO_MH2_xCOLD_Bertemes+2018.eps}%{balmerDec_LumCO_MHII_correlation_corretta.eps}%         balmerDec_LumCO_MHII_correlation.eps}
  \caption[L$_{CO}$ and M$_{H2}$ plotted against H$\alpha$/H$\beta$.]
        {{{\it{Left panel}}: L$_{CO}$ versus H$\alpha$/H$\beta$ for a sample of 333 CO(1-0) detected xCOLD-GASS galaxies. The points are color-coded as a function of the BPT classification: SF systems (blue), composite SF and AGN systems (green), AGN (magenta), unclassifiable objects (low SNR emission lines, grey). {\it{Right panel}}: as in the left panel only for composite, AGN and unclassifiable systems. In both panels $\rho_S$ indicates the value of the Spearman correlation coefficient.}}
 % {L$_{CO}$ plotted against H$\alpha$/H$\beta$ for total ($333$) CO(1-0) detected galaxy on the xCOLD GASS sample, color code according to the BPT classes: blue, green, magenta and gray dots indicate SF, composite, AGN and unclassified galaxies, respectively. The correlation is significant for the whole sample ({\it{left panel}}), increases for the SF sample and disappears if the non SF galaxies are considered ({\it{right panel}}), as indicated by the Spearman correlation coefficients ($\rho_{S}$).}
     
              \label{correlazioneLCO_BD_BPT}%
   \end{figure}
   
   %_____________________________________tabella con le informazioni sul fit__________________________
\begin{table*}  % valori della tabella in valori_fit_Lco_Mh2_BD.dat
\small
\centering
\caption{ {. L$_{CO}$-BD and M$_{H2}$-BD best fit coefficients, for all the SF and the less inclined (i$\leq$65$^{\circ}$) SF galaxy samples,  obtained by using the prescription of }\citealt{Cappellari+2013}. 
%The parameters a, b, $\sigma$, $\Delta_{obs}$ and N indicate the
%slope, intercept, internal scatter, observed RMS of the residuals around the ordinate axis of the best-fit and the number of galaxies in each sub-sample, respectively.
{$\sigma$, $\Delta_{obs}$ and N indicate the internal scatter, observed RMS of the residuals around the ordinate axis of the best-fit and the number of galaxies in each sub-sample, respectively.}}
\begin{tabular}{c|c|c|c||c|c|c|c|c|c|c|c|c}

%      &   &  \multicolumn{0}{c|}{$\log(L_{CO}) = a \times \big(\log(BD) - x_{0}\big) +  b$} &   & \multicolumn{4}{c}{ $\log(M_{H2}) = c \times \log(BD) +  d$} \\
 &   &\multicolumn{2}{|c|}{ $\log(L_{CO}) = a \times \big(\log(BD) - x_{0}\big) +  b$} & \multicolumn{7}{r|}{ $\log(M_{H2}) = c \times  \big(\log(BD) - x_{0}\big) +  d$}\\
 \hline
  \hline
                        & a                 & b                &  $\sigma$      & $\Delta_{obs}$ & & & c               & d                &  $\sigma$         & $\Delta_{obs}$  & &N  \\
       \hline
SF                     &6.53 $\pm$0.44&8.33$\pm$0.03 &  0.41 $\pm$ 0.03   &  0.46  &  &  & 5.59$\pm$0.36     & 8.94$\pm$0.03    &  0.30$\pm$0.02    & 0.39 & & 198  \\
i$\leq$65$^{\circ}$    &8.52 $\pm$0.55&8.50$\pm$0.04 &  0.33 $\pm$ 0.03   &  0.41  &  &  & 6.5$\pm$0.46      & 9.07$\pm$0.03    &  0.24$\pm$0.03    & 0.35 & & 127  \\

 \hline
\end{tabular}

\label{tab:fit1}
\end{table*}
\normalsize
%_________________________________________________________________________________________________________
%______________________________________________ 

% Singoli BDecrements Brinchmann correlazione LCO e MHII           %figura fatta con
%plot_correlation_Lco_BD_final_colorcode_NuovoFit.pro
%plot_correlation_Mh2_BD_final_colorcode_NuovoFit.pro (rosa fit brutto) e 
%plot_correlation_MH2_BD_final_colorcode.pro (rosa fit bruttino)
%plot_bootstrap_correlazione_BalmerDec_LuminositaCO_SF.pro e plot_bootstrap_correlazione_BalmerDec_MH2_SF.pro
%plot_correlation_BalmerDec_Mmolecolare_COLD+bertemes.pro.pro + plot_correlation_BalmerDec_LuminositaCO_COLD+bertemes.pro.pro + knote plot_BalmerDecrement
   \begin{figure*}    
 %  \centering
   \includegraphics[angle=-90,width=\hsize , trim={0.02cm 0 0 0}, clip]{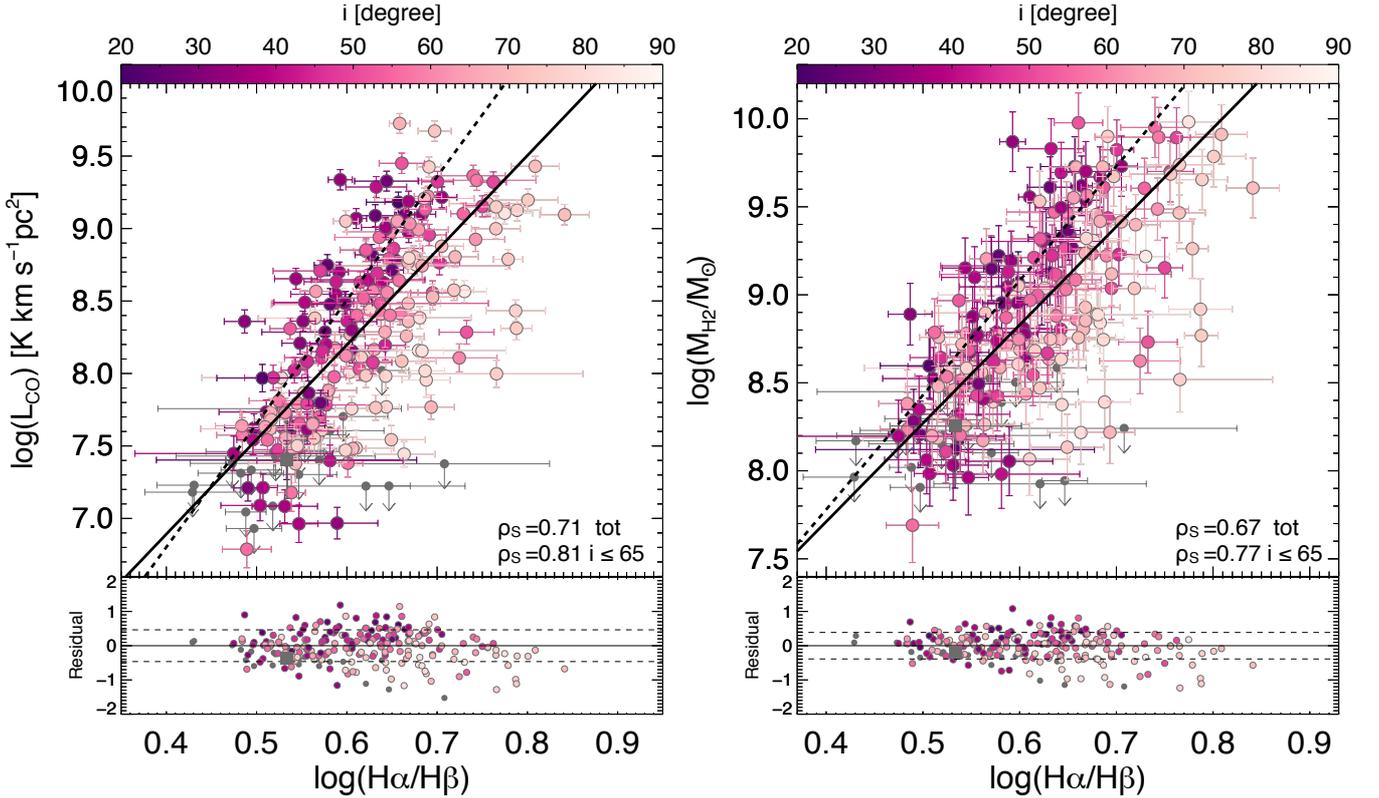}%{BD_LCO_MH2_xCOLD_fit_colorcode_inc.eps}%{BD_LCO_MH2_xCOLD_fit_bootstrapping_nuovissima}%{BD_LCO_MH2_xCOLD_fit_bootstrapping_nuova.eps}%{BD_LCO_MH2_xCOLD_fit_bootstrapping_nuova_resColor.eps}%{BD_LCO_MH2_xCOLD_fit_bootstrapping.eps}%{BD_LCO_MH2_xCOLD_Bertemes+2018_SNR10.eps}%{BD_LCO_MH2_xCOLD_Bertemes+2018.eps}%{balmerDec_LumCO_MHII_correlation_corretta.eps}%         balmerDec_LumCO_MHII_correlation.eps}
   \caption[L$_{CO}$ and M$_{H2}$ plotted against H$\alpha$/H$\beta$.]
 {{{\it{Upper panels}}: L$_{CO}$ (\textit{left}) and M$_{H2}$ (\textit{right}) versus H$\alpha$/H$\beta$ for 222 SF xCOLD GASS galaxies, color-coded as a function of the disk inclination. L$_{CO}$ luminosity upper limits and their average value are indicated with gray points and square, respectively. The best fit power law is shown for the whole sample (solid line) and for the subsample at low disk inclination (i$\leq$ 65$^{\circ}$, dashed line). The Spearman coefficient, $\rho_S$, is reported in both panels for the whole sample and the low inclination subsample. {\it{Lower panels}}: residuals as a function of H$\alpha$/H$\beta$ with respect to the best fit power law obtained for the whole sample.}}
 %{L$_{CO}$ (\textit{left}) and M$_{H2}$ (\textit{right}) plotted against H$\alpha$/H$\beta$ for $222$ local SF galaxies, $198$ with detected CO(1-0) line flux and $24$ upper limits from xCOLD GASS survey, big colored and small gray circles, respectively. 
%  Both the L$_{CO}$ and M$_{H2}$ clearly show a strong correlation with the H$\alpha$/H$\beta$ as indicate by the Spearman correlation coefficient ($\rho$) in the figure. The most inclined systems tend to exhibit extremely high values of the Balmer Decrement compare to the less inclined galaxies (as shows by the color coding). The disk self-obscuration leads to large values of BD with respect to the actual amount of dust that is connected to the molecular clouds. The Spearman correlation coefficients increase if only the less inclined galaxies (i$\leq$ 65$^{\circ}$) are taken into account.The gray square is the average value of the L$_{CO}$ and M$_{H2}$ upper limits. The black solid and dashed lines are the best linear log-log fits obtained by fitting all the $198$ galaxies with detected CO(1-0) line flux and the low inclined galaxies. 
%The residuals are shown as a function of H$\alpha$/H$\beta$ in the lower panels.}
              \label{correlazioneLCO_MH2_BD}%
   \end{figure*}
%______________________________________________ 
For all galaxies in the xCOLD GASS sample the L$_{CO}$ is converted to the total molecular gas mass, M$_{H2}$ by using the empirical relation: $  M_{H2}= \alpha_{CO} \times  L_{CO}  $, where $\alpha_{CO}$ is the CO$-$to$-$H$_2$ conversion factor (\citealp{Dickman+1986,ObreschkowRawlings2009} and \citealp{KennicuttEvans2012} for a review) calibrated in \citealp{Accurso+2017}.

The xCOLD GASS catalog also provides several global galaxy properties as: stellar mass (M${\star}$), star formation rate (SFR), classification of the galaxies according to main ionization mechanism (star-forming, composite, AGN and unclassified galaxies according to the BPT diagram, \citealp{BPT}), galaxy size ($r_{50}$, the radius encapsulating 50\% of the Petrosian r-band flux), gas-phase metallicity $12 + \log$ O/H (obtained from the N[II]/H$\alpha$ and O[III]/H$\beta$ ratio and the calibration of \citealp{PettiniPagel2004}) and {concentration index\footnote{C$_{index} = r_{90}$/$r_{50}$ is defined as the ratio of the r-band Petrosian radii encompassing 90\% and 50\% of the flux, which can be used as a proxy for the bulge-to-disk ratio, see e.g. \citealp{Weinmann+2009}}.}  
%(from WISE + GALEX when detected in both data sets else from SED fitting, see \citealt{Janowiecki2017}). See \cite{Saintonge+2011,Saintonge+2017} for a complete description of the sample selection, survey strategy and associated data. 

To retrieve an accurate value of the galaxy disk inclination we use the measures provided by the morphology catalogue of \cite{Simard+11} when available, and the xCOLD GASS sample otherwise. All the 532 xCOLD GASS galaxies are cross matched with SDSS DR7 MPA-JHU emission line catalog\footnote{https://wwwmpa.mpa-garching.mpg.de/SDSS/DR7/raw$\_$data.html}, witch provides the measurements of H$\alpha$ and H$\beta$ fluxes and related uncertainties necessary to evaluate the Balmer Decrement (BD). 
%As explained in Chapter 2, the emission line fluxes are taken from the revised version of the MPA-MJU catalog\footnote{http://www.mpa-garching.mpg.de/SDSS/DR7/} adapted to the DR7 data release. The emission line fluxes are estimated after continuum removal. 
%The spectral continuum is removed by fitting each galaxy spectrum with a set of sets of spectral energy distribution templates (from Charlot \& Bruzual in prep. CB08 models) at the SDSS spectral resolution. 
%After subtracting the best-fitting stellar population model of the continuum, any remaining residual is removed with a sliding 200 pixel median and fit the nebular emission lines. To recover also very weak nebular features, all emission lines are fitted simultaneously with Gaussians, requiring that all of the Balmer lines (H$\alpha$, H$\beta$, H$\delta$, and H$\gamma$) have the same line width and velocity offset (see for details \citealt{Tremonti+04}).
As discussed on the MPA-JHU website, and well explained by \cite{Groves+2012} and \cite{Brinchmann2013}, the listed errors of the MPA/JHU catalogue are formal, and likely underestimated. Following the approach of Brinchmann and collaborators, we multiplied the uncertainty of the emission lines by a correction factor ($f=2.473$ and $f=1.882$ for the H$\alpha$ and H$\beta$ line respectively), to take into account continuum subtraction errors.

All xCOLD GASS galaxies have a mach in the MPA/JHU catalogue and so a measure of the BD. In Fig. \ref{FigPlaneBalmerSingoli} we illustrate the distribution of our sample in the SFR-M$\star$ plane. The points are color-coded as a function of the BD signal-to-noise-ratio, SNR(H$\alpha$/H$\beta$). As expected, the BD is well detected in galaxies on and above the Main Sequence (MS) of star forming galaxies but it shows a progressively lower SNR at larger distances from the MS towards the quiescence region. Finally, by cross matching the xCOLD GASS sample with the xGASS sample provided by \cite{Catinella+2018} we obtained a subsample of $253$, with a estimate of atomic gas mass (M$_{HI}$), M$_{H2}$ and BD.

%, where the detection level of the emission lines is quite low.

%In Fig. \ref{FigPlaneBalmerSingoli} we illustrate the distribution of our sample in the SFR-M$\star$ plane, color-coded with their H$\beta$ signal-to-noise ratio (S/N). 
%The H$\beta$ line is well detected in galaxies with high SFR, above and in MS of star forming galaxies, H$\beta$ S/N $>10$, while it is purely observed at low SFRs, H$\beta$ S/N $<10$. The same trend is obtained for the CO line detection (\citealp{Saintonge+2017}, Fig. 7 right panel).

%For the purposes of this study, we select only the galaxies with well detected emission line fluxes. After a selection in signal to noise in the Balmer decrement (SNR$>10$), we end up with a final sample of $256$ galaxies, $195$ from the XCOLD GASS sample and $61$ from the \cite{Bertemes+2018} catalogue. 
%$133$ galaxies with low H$\beta$ flux, (S/N$<10$) are rejected to minimize the error in the final Balmer Decrement.

\section{The Balmer decrement-L$_{CO}$ and M$_{H2}$ correlation}
%_____________________________________tabella con le informazioni sul fit__________________________
\begin{table*}  % valori della tabella in valori_fit_Lco_Mh2_BD.dat
\small
\centering
\caption{ {L$_{CO}$-BD-i and M$_{H2}$-BD-i  best fit coefficients, for all the SF galaxies, obtained by using the prescription of }\citealt{Cappellari+2013}. 
%The parameters a, b, c, $\sigma$, $\Delta_{obs}$ and N indicate the
%slope, intercept, internal scatter, observed RMS of the residuals around the ordinate axis of the best-fit and the number of galaxies in each sub-sample, respectively.
{$\sigma$, $\Delta_{obs}$ and N indicate the internal scatter, observed RMS of the residuals around the ordinate axis of the best-fit and the number of galaxies, respectively.}}
\begin{tabular}{c|c|c|c||c|c|c|c|c|c|c|c|c}

 & &\multicolumn{4}{|c|}{ $\log(L_{CO})= a + b \times \big(\log(BD)-\log(BD)_{0})\big) + c \times \Big( i-~i_{0}\big)$} & \multicolumn{6}{r|}{ $\log(M_{H2})= d + f \times \big(\log(BD)-\log(BD)_{0})\big) + g \times \Big( i- i_{0}\big)$}\\
 \hline
 \hline
              & a                 & b        &     c   &  $\sigma$      & $\Delta_{obs}$ &  & d              & f &g                &  $\sigma$         & $\Delta_{obs}$  &N  \\
       \hline
SF                     &8.32 $\pm$0.03&  7.59$\pm$0.43 & -0.013 $\pm$ 0.002      &  0.36 $\pm$ 0.03   &  0.43  &           & 8.93$\pm$0.03     & 5.86$\pm$0.35  & -0.010$\pm$0.002   &  0.26$\pm$0.03    & 0.36  & 198  \\
%i$\leq$65$^{\circ}$    &8.52 $\pm$0.55&8.50$\pm$0.04 &  0.33 $\pm$ 0.03   &  0.41  &  &  & 6.5$\pm$0.46      & 9.07$\pm$0.03    &  0.24$\pm$0.03    & 0.35 & & 127  \\
 \hline
\end{tabular}

\label{tab:fit2}
\end{table*}
\normalsize
%_________________________________________________________________________________________________________
%______________________________________________ figura fatta con  Lco_BD_inclianzione_fit_errori.pro (forse)
   \begin{figure*}%[!t]    
   \centering
   \includegraphics[angle=-90,width=\hsize, trim={0.02cm 0 0 0}, clip ]{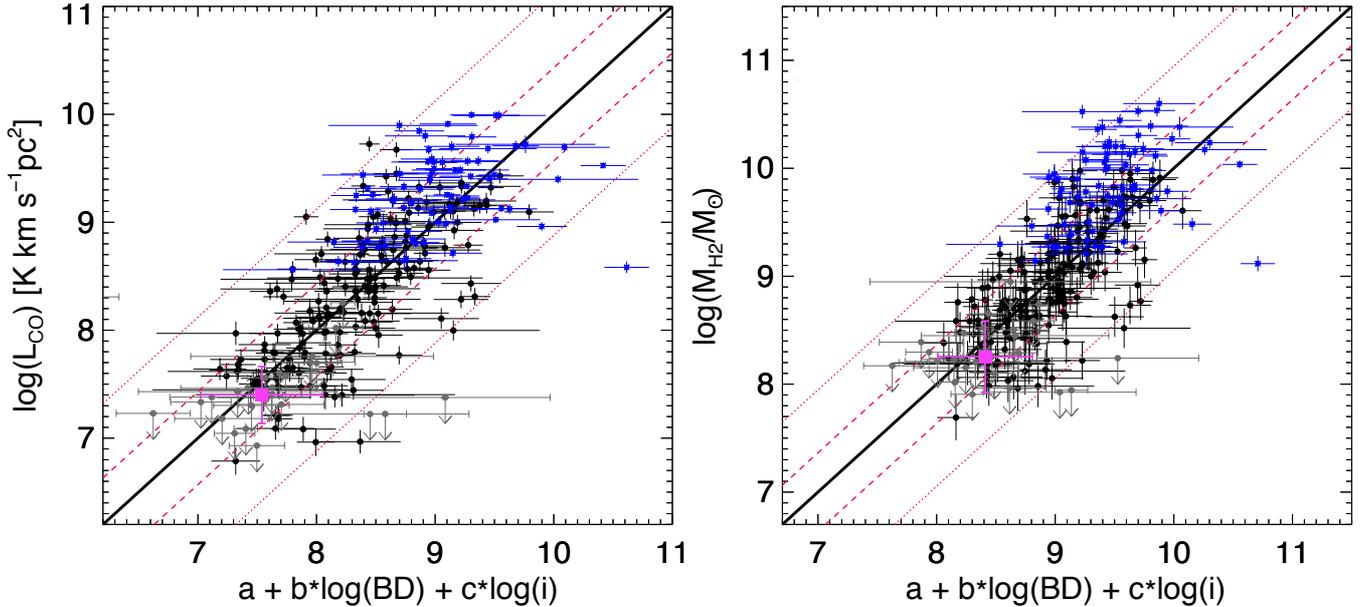}%{balmerDec_LumC0_MH2_i_plane_XCOLDGASS_Bertemes2018_new.eps}%{balmerDec_LumC0_MH2_i_plane_XCOLDGASS_Bertemes2018.eps}%{balmerDec_LumC0_i_plane_XCOLDGASS_Bertemes2018.eps}%{CopiadibalmerDec_inclinazione_LumC0_correlation_XCOLDGASS_bootstrapping_SF_conUpperLimits_Bertemes.pdf}%CopiabalmerDec_inclinazione_LumC0_correlation_XCOLDGASS_bootstrapping_SF_conUpperLimits.pdf
   \caption[L$_{CO}$ plotted against H$\alpha$/H$\beta$ and disk inclination.]
  {{L$_{CO}$ and M$_{H2}$ versus} H$\alpha$/H$\beta$ {(BD)} corrected for the disk inclination (i) {for for the same sample as in Fig. 3. Upper limits are indicated by gray points.} The magenta square is the average value of the upper limits.
{The red lines mark the 1$\sigma$ (dashed) and 2.6$\sigma$ (dotted) regions.} The errors bars are the projection of the observational errors. The blue stars represent the 78 starburst galaxies taken from the \cite{Bertemes+2018} catalogue.}
           
              \label{Lco_BD_inclinazione_piano_berte}%
   \end{figure*}
%______________________________________________ 

%______________________________________________ 
%   \begin{figure}%[!t]    
%   \centering
%   \includegraphics[angle=0,width=\hsize ]{residuiLco_BD_new.eps}%{residuiLco_BD.eps}%         balmerDec_LumCO_MHII_correlation.eps}
%   \caption[residual bias]
%  {Residuals around the best L$_{CO}$-BD relation. The residuals are defined as the difference between the CO luminosity observed on our sample and the CO luminosity derived from the $BD$-L$_{CO}$ best fit relation. 
%  {\it{Upper left panel}}: relation between the residuals and the ratio between the galaxy observed r50 and the fiber SDSS radius of 1.5 arcsec. {\it{Upper right panel}}: relation between residuals and galaxy inclination as derived from \cite{Catinella+2012}. {\it{Middle left panel}}: relation between residuals and gas-phase metallicity $12 + \log$ O/H adopting the \cite{PettiniPagel2004} relations. {\it{Middle right panel}}: concentration index, C$_{index} = r90/r50$, defined as the ratio of the r-band Petrosian radii encompassing 90\% and 50\% of the flux, which can be used as a proxy for the bulge-to-disk ratio.
%  {\it{Bottom left panel}}: relation between the residuals and the stellar mass.  {\it{Bottom right panel}}: relation between the residuals and the star formation rate taken from the MPA-JHU database. In all panels the magenta lines show the 1$\sigma$ scatter of the L$_{CO}$-BD relation and the zero value, dotted and continuous lines, respectively.}
%              \label{residuals}%
%   \end{figure}
%______________________________________________ 

First, we study the correlation between the luminosities of the ~CO(1-0) line, L$_{CO}$, versus the Balmer decrement, (BD$=$H$\alpha$/H$\beta$) in the xCOLD GASS  subsample with a detected CO(1-0) emission (333 galaxies).
%in all the 333 galaxies with a detected CO emission line and in galaxies that are classified according the BPT diagram as: unclassified (UnClass), star-forming (SF), composite (AGN+SF) and AGN dominated systems (AGN).
%In Fig. \ref{correlazioneLCO_MH2_BD} L$_{CO}$ and BD are compared in the 333 galaxies with a detected CO(1-0) emission line (left panel), star-forming (SF) galaxies (middle panel), unclassified, composite and AGN dominated (UnClass, AGN+SF and AGN) systems (right panel).
The correlation is significant when the whole subsample is considered, with a Spearman rank coefficient \citep{Spearman1904} $\rho_{S}$ of 0.5 and a 99\% probability of correlation (left panel of Fig.\ref{correlazioneLCO_BD_BPT}). 
%However, the scatter of the correlation decreases and its significance increases substantially 
The significance of the correlation increases substantially ($\rho_{S}=0.71$) and the scatter decreases when only pure SF galaxies are considered. Instead, unclassified, composite SF/AGN and AGN dominated galaxies exhibit a much larger scatter and a poor correlation  ($\rho_{S}=0.26$, right panel of Fig.\ref{correlazioneLCO_BD_BPT}). 
Such lack of correlation in the non SF galaxies is not surprising. %Indeed, is well established that the optical emissions (H$\alpha$ and H$\beta$ and so the BD) are very noisy on unclassified galaxies, while in the AGN and composite systems, they are probably affected by the photons originates in the narrow line regions which are not directly connected with the presence of cold molecular gas.
Indeed, unclassified galaxies tend to be located in the quiescence region, where the BD SNR is poor (\ref{FigPlaneBalmerSingoli}), In addition, in AGN hosts and composite systems the optical emission
is probably affected by the photons originated in the narrow line regions which are not directly connected with the presence of cold molecular gas.
%is affected by the AGN contribution.

We limit the study of the correlation to the 226 pure SF galaxies, of which 201 galaxies with detected CO(1-0) emission and {25 L$_{CO}$ upper limits. %To avoid spurious results
 %we remove from the final SF sample 4 galaxies which show complex morphology (merging like systems) and/or uncorrected BD measurement.}
%To quantify the correlation, 
To avoid spurious results due to the presence of galaxies with complex morphology (merging like systems) and/or uncorrected BD measurement, we remove 4 object from the SF sample.% CO detected sample (201-3=198) and 1 galaxy from the SF CO undetected sample (25-1=24). 
 The final SF sample consist of 222 SF galaxies,  198 with detected CO emission and 24 upper limits ( see Fig. \ref{correlazioneLCO_MH2_BD} ).}
The data are fitted with a power law of the form: $\log(L_{CO}) = a \times \big(\log(BD) - x_{0}\big) +  b$ by using the LTS$\_$LINEFIT code presented by \cite{Cappellari+2013},  which allows for errors in both variables and intrinsic scatter (defined as the dispersion of the data along the y-direction due to the physical values rather than to measurement errors, see \citealp{Tremaine+2002, Cappellari+2013} for more details). {To minimize the uncertainty on the intercept (b), we use a reference value close to the middle of the x values, x$_{0}=0.62$.}
The best fit power law is showed in Fig.  \ref{correlazioneLCO_MH2_BD}, left panel. The best fit coefficients, internal scatter and observed RMS along the y-coordinate are reported in Tab. \ref{tab:fit1}, left part.
%\begin{equation}
%    \log(L_{CO}) = 9.66{\pm 0.57} \times \log(BD) +  2.37{\pm 0.35} \qquad .
%\end{equation}
%The errors on the best-fit linear relation are determined by bootstrapping analysis.
%%%%%  DA DECIDERE MA NN SEMBRANO ESSERCI GROSSE DIFFERENZE IN TERMINI DI ERRORI TRA ERRORI DAL FIT DI MICHELE E DA 1000 RIPETIZIONI
%To estimate the error of the best fit parameters, we generate 1000 random catalogues by scattering each point according to 1$\sigma$ error in both L$_{CO}$ and BD. The fit is reiterate in each catalogue as explained above. Finally, the error associated to the fit parameters is computed as the dispersion of 1000 fit results. 
The fitting procedure does not include the L$_{CO}$ upper limits. However, we point out that such upper limits and their mean value (grey symbols, in left panel of Fig. \ref{correlazioneLCO_MH2_BD}) follow the average trend.

By adopting a constant conversion factor (e.g. $\alpha^{MW}_{CO} = 4.3$ ~M$_{\odot}$/K km s$^{-1}$ pc$^{2}$ for the Milky Way, see the discussion in \citealp{Bolatto+2013}) this correlation can be easily translated to a M$_{H2}$-BD correlation of the form: $\log(L_{CO})+\log(\alpha^{MW}_{CO})=\log(M_{H2})$.
Differently, if a non-constant conversion factor is assumed, the shape of the relation could be different. 
%slightly changes (ma dovresti dire se sono consistenti o meno entro 1 sigma) a naso direi di si.
As an example, in the right panel of Fig. \ref{correlazioneLCO_MH2_BD} we report the molecular mass, M$_{H2}$-BD relation obtained by using the M$_{H2}$ obtained with the $\alpha_{CO}$ proposed by \citealp{Accurso+2017}, witch depends primarily on gas metallicity and secondarily on the offset from the star-forming main sequence.
%Also in this case, we find a positive correlation between BD and M$_{H2}$, confirmed by a Spearman correlation coefficient of $\rho=0.67$ and with a probability of correlation above $99\%$.
%The best fit parameters, intrinsic scatter and observed RMS are reported on Tab. 1, right part.

%We find that both the observed RMS and internal scatter are $\Delta_{obs}$=0.46 dex and $\sigma$=0.41 $\pm$ 0.03, in the L$_{CO}$, and $\Delta_{obs}$=0.39 dex and $\sigma$=0.30 $\pm$ 0.025 in the $M_{H2}$ case.%, with a clear deviation towards high value of BD, above $\sim 5$ (bottom panels of Fig. \ref{correlazioneLCO_MH2_BD}). 
In order to understand what are the possible sources of scatter we first study our L$_{CO}$-BD and M$_{H2}$-BD correlations against the galactic disk inclination. % measures provided by the morphology catalog of \cite{Simard+11} or the inclination values provided by the xCOLD GASS sample when they are not available. 
As expected, systems with the highest inclination tend to exhibit extremely high values of the BD compared to the less inclined galaxies (as shown by the colour coding on Fig. \ref{correlazioneLCO_MH2_BD}). This is likely due to the fact that in edge-on systems, the H$\alpha$ and H$\beta$ photons need to pass through the entire disk before escaping the galaxy, with a large probability of being absorbed by the disk itself. The disk self-obscuration leads to large values of BD with respect to the actual amount of dust that is linked to the molecular clouds. 

To take into account this effect, we test two different approaches.
%Indeed, we point out that the Spearman correlation coefficient of the L$_{CO}$-BD relation increases from $0.71$ to $0.81$ (and from $0.67$ to $0.77$ in the M$_{H2}$ case) if only galaxies with inclination smaller that $65^\circ$ are considered.
First, we reduce our sample on galaxies with i$\leq 65^\circ$. 
In this case, the Spearman correlation coefficient increases in both L$_{CO}$-BD and M$_{H2}$-BD correlations. Similarly the intrinsic and the observed scatter decreases respectively to $\sim 0.3$ and $\sim 0.4$ dex in L$_{CO}$ and to $\sim 0.24$ and $\sim 0.4$ dex in M$_{H2}$ case.
%We find that the Spearman correlation coefficients of the L$_{CO}$-BD and M$_{H2}$-BD relation increase (from $0.71$ to $0.81$ and from $0.67$ to $0.77$ in the L$_{CO}$ and M$_{H2}$ case, respectively). By repeating the log linear fits for this subsample of less inclined galaxies, we find 
%The results of the first approach are showed with the dashed lines in Fig. \ref{correlazioneLCO_MH2_BD}, and  with the fit parameters reported in the second line of Tab. \ref{tab:fit}. 
%In both the L$_{CO}$ and M$_{H2}$ cases, we find
%a steeper and tighter relation with a clear decrease of the intrinsic and observed scatter, $\sigma$ goes from $0.41$ to $0.33$ and $\Delta$ from $0.46$ to $0.41$ dex in L$_{CO}$ relation ($\sigma$ from $0.41$ to $0.33$ and $\Delta$ from $ $ to $ $ dex in the M$_{H2}$ case). 
The fit parameters are reported on Tab. 1. As an alternative approach, we parametrize the L$_{CO}$ {as a function of BD and the galaxy inclination using the plane fitting prescription proposed by \cite{Cappellari+2013}:~$\log(L_{CO})= a + b \times \big(\log(BD)-\log(BD)_{0})\big) + c \times \Big( i-~i_{0}\big)$, 
%and  $\log(M_{H2})= a + b \times \big(\log(BD)-\log(BD)_{0})\big) + c \times \Big( i- i_{0}\big)$,
where $\log(BD)_{0}= 0.62$ and $i_{0}=56^\circ$ are the adopted reference values. We use the same form for the M$_{H2}$-BD relation. The best fit values are reported in Tab.\ref{tab:fit2} for both the L$_{CO}$ and M$_{H2}$.}  %, with $Y=$ to L$_{CO}$ or M$_{H2}$. The best fit parameters are reported in Fig. \ref{Lco_BD_inclinazione_piano_berte}.} %Also in this case the errors are estimated by using the N$=1000$ catalogues discussed above.
Also in this case, the correlations are tighter, with a reduction of the intrinsic and observed scatter of 0.36$\pm$ 0.03 and 0.43 dex in L$_{CO}$ and 0.26$\pm$ 0.03 and 0.36 dex on the M$_{H2}$ case.

{As reported from several previous studies, the gas-to-dust ratio clearly depends on gas metallicity (e.g. \citealp{Draine2007,Galametz+2011} ). However, we notice that our correlation is stable across our gas metallicity range, $8.45<12 + \log O/H<8.8$.}
%I see another clear issue with the conversion to Mgas in the fact that the gas/dust ratio depends on metallicity as well hence on the dust mass, from which one might expect non-linear effects between BD and Mgas.  All these aspects should be clearly discussed. 
To test the L$_{CO}$-BD and the L$_{CO}$-BD-i correlations against biases, we study the distribution of the residuals against several parameters such as the galaxy size, concentration index, stellar mass and SFR. %The results of such analysis are shown in Fig. \ref{residuals}, in none of cases we see a correlation, as confirmed also by the Spearman test.
No correlations are observed in both the L$_{CO}$-BD and the L$_{CO}$-BD-i case, between the residuals and the other parameters considered, as confirmed by the Spearman test.
%{In particular we notice that the correlation is stable across our gas metallicity range, $8.45<12 + \log O/H<8.8$.}

Finally, to test the reliability of the correlation, we overplot in Fig. \ref{Lco_BD_inclinazione_piano_berte} the {\it{Herschel}} SPIRE selected sample of \cite{Bertemes+2018} in the SDSS Stripe82 area. The shallow {\it{Herschel}}
Stripe82 survey leads to the selection of the dustiest local objects. \cite{Bertemes+2018} provide a measure of L$_{CO}$ for a subsample of WISE and SPIRE simultaneously detected systems with Alma CO follow-up. %We calculate the M$_{H2}$ following the prescription of \cite{Saintonge+2017}.
Such objects, which sample the highest L$_{CO}$ and M$_{H2}$ end, remarkably follow the correlation given by the xCOLD GASS sample.

\section{The Balmer decrement-M$_{HI}$ correlation}
% correlazione     MHI          
%figura fatta con correlazione_BalmerDecrement_MHI.pro + knote
   \begin{figure}    
   \centering
   \includegraphics[angle=90,width=\hsize ]{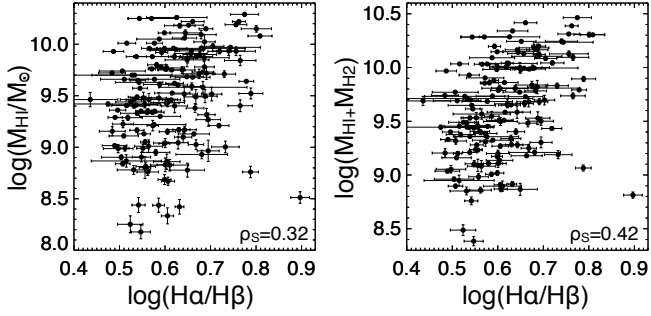}
   \caption[M$_{HI}$ (left) and total gas mass (M$_{H2}$+M$_{HI}$, right) versus H$\alpha$/H$\beta$.]
   {{M$_{HI}$ (left) and total gas mass (M$_{H2}$+M$_{HI}$, right) versus H$\alpha$/H$\beta$ for 158 SF galaxies with molecular and atomic estimates.
  The Spearman correlation coefficient $\rho_S$ is indicated in both panels. The errors in the atomic mass are evaluated by taking into account the errors in the HI flux.}}
              \label{FigMHI_MH2_BD}%
   \end{figure}
%______________________________________________ 

For a subsample of $158$ SF galaxies with both atomic and molecular mass, we consider also the relation between Balmer decrement and atomic gas mass (M$_{HI}$) taken from the xGASS catalogue (see \citealt{Catinella+2018}). First, one should consider that the molecular mass itself does not correlate significantly with the HI gas mass, as found in \cite{Catinella2013,Catinella+2018} %we observe a very poor and noisy relation between $M_{H2}$ and $M_{HI}$. The best fit relation lies well above the 1 to 1 relation because the atomic gas mass is much larger than the molecular gas mass. 
This is due to the fact that the molecular gas mass is located mainly in the star forming region of the galaxy stellar disk, while the HI disk is much more extended \citep{Catinella+2018}. 
Similarly the BD does not correlate strongly with the HI mass as shown in the left panel of Fig. \ref{FigMHI_MH2_BD}. In the same way, the total cold gas mass, given by the sum of the HI and H2 gas mass, shows only a poorly significant correlation with the BD (right panel of Fig. \ref{FigMHI_MH2_BD}), as the atomic phase is dominating over the molecular gas.

\section{Conclusions}\label{disc_conclusions}
We find a tight correlation between the dust screen surrounding the HII regions, traced by the Balmer Decrement (BD) and the cold molecular gas (M$_{H2}$) traced by the CO(1-0) luminosity (L$_{CO}$) in a sample of $222$ local star-forming  galaxies taken from the xCOLD GASS survey. In AGN, composite and unclassified galaxies the correlation is not visible due to the contamination by the nuclear region or the very low SNR of the Balmer emission lines.

As expected, the galaxy inclination leads to the observation of extremely high value of BD in edge-on galaxies.
Once corrected for the disk self-obscuration in such highly inclined galaxies, the BD can be used as a very powerful proxy of the L$_{CO}$ and the M$_{H2}$ in local star forming galaxies, with a scatter of $\sim$0.3 dex. 
We test the correlation against possible biases {induced by the physical properties of our sample, as} the galaxy size, mass, morphology, star formation activity and gas metallicity but we do not find any dependence of the residuals on the considered parameters.
The retrieved relation also matches the L$_{CO}$-BD distribution of the dustiest local objects, identified in the Herschel Stripe82 survey (\citealp{Bertemes+2018}).
{We highlight the fact that our relation is calibrated on local massive (M$\star > 10^{9}$ M$_{\odot}$) and metal rich galaxies ($12 + \log O/H > 8.45 $), any future application outside of its range of validity should be take with caution. }
We do not find a significant correlation if the atomic gas phase is taken into account. This is likely due to the fact that the region traced by the BD, the stellar disk, is much smaller than the HI disk.

%I see another clear issue with the conversion to Mgas in the fact that the gas/dust ratio depends on metallicity as well hence on the dust mass, from which one might expect non-linear effects between BD and Mgas.{We emphasize the fact that the gas metallicity range covered in our sample is very small, , so we cannot explore the variation of the gas-to-dust whit the metallicity observed in previous studies.} a relation calibrated on local data outside of its range of validity – a similar caution would be valid for an application at high redshift as well

\section*{Acknowledgements}
AC thanks Federico Lelli for precious suggestions and stimulating discussion. AC is grateful to Luca Cortese, Roberto Maiolino and Asa F. L. Bluck for the interesting suggestions and Barbara Catinella for the clarification concerning the COLD GASS dataset. 
This research was supported by the DFG cluster of excellence "Origin and Structure of the Universe''.
%This work made use of the xCOLD GASS IRAM-30m legacy survey \citep{Saintonge+2011,Saintonge+2017} and COLD GASS Arecibo SDSS Survey \citep{Catinella2010,Catinella+2018}.
The authors acknowledge all the members of the xCOLD GASS \citep{Saintonge+2011,Saintonge+2017} and COLD GASS \citep{Catinella2010,Catinella+2018} projects for providing the data.

\bibliographystyle{mnras}
\bibliography{alice} % if your bibtex file is called example.bib

% Don't change these lines
\bsp	% typesetting comment
\label{lastpage}
\end{document}